\pdfoutput=1
\def\year{2022}\relax
\documentclass[letterpaper]{article} %

\usepackage{aaai22}  %
\usepackage{times}  %
\usepackage{helvet}  %
\usepackage{courier}  %
\usepackage[hyphens]{url}  %
\usepackage{graphicx,epstopdf} %
\usepackage{natbib}  %
\usepackage{caption} %
\DeclareCaptionStyle{ruled}{labelfont=normalfont,labelsep=colon,strut=off} %
\usepackage{algorithm}
\usepackage{algorithmic}

\usepackage{newfloat}
\usepackage{listings}
\floatstyle{ruled}
\newfloat{listing}{tb}{lst}{}
\floatname{listing}{Listing}
\title{DiPS: Differentiable Policy for Sketching in Recommender Systems}

\author {
    Aritra Ghosh\textsuperscript{\rm 1},
    Saayan Mitra\textsuperscript{\rm 2},
    Andrew Lan\textsuperscript{\rm 1} \\
}
\affiliations {
    \textsuperscript{\rm 1}University of Massachusetts Amherst \\
    \textsuperscript{\rm 2}Adobe Research \\
    arighosh@cs.umass.edu, smitra@adobe.com, andrewlan@cs.umass.edu
}

\usepackage[utf8]{inputenc} %
\usepackage[T1]{fontenc}    %
\usepackage{hyperref}       %
\usepackage{url}            %
\usepackage{booktabs}       %
\usepackage{amsfonts}       %
\usepackage{nicefrac}       %
\usepackage{microtype}      %
\usepackage{xcolor}         %
\usepackage{amssymb}
\usepackage{algorithm,algorithmic,amsmath}
\usepackage{graphicx}
\usepackage{xcolor}
\usepackage{amsmath,bm}
\usepackage{graphicx}
\usepackage{textcomp}
\usepackage{xcolor}
\usepackage{bbm}
\usepackage{url}
\usepackage{wrapfig}
\usepackage{amsthm,multirow}

\newcommand{\bx}{\mathbf{x}}

\newcommand{\bz}{\mathbf{z}}
\newcommand{\bw}{\mathbf{w}}

\newcommand{\by}{\mathbf{y}}

\newcommand{\bu}{\mathbf{u}}

\newcommand{\bv}{\mathbf{v}}

\newcommand{\be}{\mathbf{e}}

\newcommand{\tth}{\text{th}}

\newcommand{\bp}{\mathbf{p}}

\newcommand{\CR}{\mathcal{R}}

\newcommand{\CE}{\mathcal{E}}
\newcommand{\CI}{\mathcal{I}}
\newcommand{\CL}{\mathcal{L}}
\newcommand{\CS}{\mathcal{S}}

\newcommand{\BR}{\mathbb{R}}

\newcommand{\CM}{\mathcal{M}}

\newcommand{\BE}{\mathbb{E}}

\DeclareMathOperator*{\argmin}{\arg\!\min}
\begin{document}

\maketitle
\begin{abstract}
In sequential recommender system applications, it is important to develop models that can capture users' evolving interest over time to successfully recommend future items that they are likely to interact with. For users with long histories, typical models based on recurrent neural networks tend to forget important items in the distant past. Recent works have shown that storing a small sketch of past items can improve sequential recommendation tasks. However, these works all rely on static sketching policies, i.e., heuristics to select items to keep in the sketch, which are not necessarily optimal and cannot improve over time with more training data. In this paper, we propose a differentiable policy for sketching (DiPS), a framework that learns a data-driven sketching policy in an end-to-end manner together with the recommender system model to explicitly maximize recommendation quality in the future. 
We also propose an approximate estimator of the gradient for optimizing the sketching algorithm parameters that is computationally efficient. We verify the effectiveness of DiPS on real-world datasets under various practical settings and show that it requires up to $50\%$ fewer sketch items to reach the same predictive quality than existing sketching policies.
\end{abstract}

\section{Introduction}
Recommender Systems (RSs) have seen great success in matching users with items they are interested in when large-scale user-item interaction datasets are available. 
Early approaches in RS assume that a user's interest is static over time and uses a RS model to compute their latent \emph{interest} state from historical interactions and predict their ratings on future items \cite{ncf,deepmf}. %
Sequential RS (SRS) is an emerging research topic on how to effectively capture user preference changes over time.
The general idea is to keep track of a user's latent interest state \emph{over time}, e.g., using a recurrent neural network \cite{recurrent-recommender,gru,long-range-dependence}. 
Since recurrent neural networks are prone to forget interactions in the distant past, in many practical scenarios, sequential RS approaches prioritize on making decisions according to a user's recent interactions in the current \emph{session} \cite{gru4rec,gru4rec++}. 
However, these approaches are not adept at capturing a user's static interests from their history, which can also be critical to future recommendations.
Recently, \cite{streaming-hardest,streaming-reservoir} found that storing a small set of historical items is highly beneficial in session-based SRS. 
Therefore, the sketching policy, i.e., how to select \emph{which} items to keep in the sketch, is key to the effectiveness of SRS approaches.

The sketching policy plays an important role in real-world SRS applications to reduce memory consumption and has been extensively studied in the context of problems such as moment finding, k-minimum value, and distinct element counting with probabilistic guarantees \cite{sketch-survey}. Various data structures can be used for sketching \cite{sketch-survey}; in this paper, we will focus on sample-based sketching as used in previous RS literature \cite{streaming-hardest,streaming-reservoir} where each sketch item is a past user-item interaction. 
This setting is related to problems such as data summarization and coreset construction where influence score and item hardness are often used to select representative samples \cite{hard-sample,influence}. 
Another related problem is active learning where the label uncertainty of future items is often used to select the next item to query \cite{active-survey}. 
A common theme for all these methods is that they define a measure of \emph{informativeness} to select the most informative item(s). 

The signature of RS applications is that items come in a \emph{streaming} fashion \cite{streaming}; thus, at each time step, we observe a single item, decide whether to store it in the sketch, and decide which item to remove from the sketch if the current item is stored.
This streaming nature is in stark contrast to the aforementioned problems where one often have access to the full set of items available to select from. 
This ``one-pass'' sketching setup is more challenging than the sketching setup in other problems. %
In practice, simple sketching strategies such as uniform reservoir sampling are often adopted in RS applications \cite{streaming-reservoir,streaming-hardest}. These approaches usually keep a reservoir of random historical items to compute the gradient required for model updates.

Although static sketching policies often work well, two major limitations hinder their further development. First, these methods  define a heuristic informativeness measure to select items to add or remove from the sketch that is \emph{not optimized} on the real objective of RS: predictive quality of the user's interaction with future items. Second, these informativeness metrics are \emph{static} and cannot exploit abundant information that can be extracted from larger and larger training datasets that are made available in recent years. Recently, there have been approaches for differentiable sample selection policy learning in data streams in other application domains \cite{bobcat}; however, these approaches are only applicable when the prediction objective is computed on a static set of known items. In contrast, in streaming RS sketching, items included in the prediction task are constantly changing.

\paragraph{Contributions.}
In this paper, we propose {\bf DiPS}, a \textbf{D}ifferentiable \textbf{P}olicy for \textbf{S}ketching framework to \emph{learn} a sketching policy that optimizes the performance on the final recommendation tasks in SRS.  %
The sketching policy is learned in an end-to-end manner together with the base RS model; at each time step, the policy takes the past sketch (of $K$ items) and the recent item (or items) and produces a new sketch (of $K$ items) for use in subsequent time steps. We make three key contributions:

First, we formulate the sketch update and recommendation tasks as a bi-level optimization problem \cite{bilevel} with a learnable sketching policy. 
In the outer-level optimization problem, we learn both the base RS model and the sketching policy %
by explicitly maximizing predictive quality on future recommendations. 
In the inner-level optimization problem, we adapt the base RS model for the user using the items in the current sketch. 
The sketching policy is learned in a fully differentiable manner with the sole objective of maximizing performance on future time steps. 

Second, we propose an approximate estimator of the true gradient of the sketching policy parameters using a separate queue module that is computationally efficient. 
Since at any time step, the sketch is dependent on all the prior decisions made by the sketching policy,
we need to back-propagate gradient to all the previous time steps.
This gradient computation requires a re-computation of the entire sketching process from the start until the current step using the \emph{current policy parameters}, which is computationally intensive.
Instead, we show that our approximation effectively alleviates these cumbersome computations.

Third, we verify the effectiveness of DiPS through extensive experiments on five real-world datasets. 
We observe that the learned sketching policy outperforms existing sketching policies using static informativeness metrics on future recommendation and prediction tasks, requiring up to $50\%$ fewer sketch items to reach the same predictive quality.  Our implementation will be publicly available at \url{https://github.com/arghosh/DiPS}.

\section{Methodology}
We now detail the DiPS method. We will start with notations and the generic sketching problem setup, followed by details on base RS models, the sketching policy, and how to efficiently learn the sketching policy. 
\subsection{Notation and Problem Setup} 
We use the notation $[N]$ to denote the set $\{1,\cdots,N\}$ and use shorthand notation $x_{1:t}$ for the set $\{x_1,\cdots,x_t\}$.
There are a total of $N$ users, indexed by $i\in [N]$ and $M$ items, indexed by $j\in [M]$. 
For notation simplicity, we will only discuss the sketching process for a single user, with a total of $T$ discrete time steps, i.e., interactions, indexed by $t$. 

We consider two commonly studied RS settings. 
In the {\bf explicit} RS setting, for a user, the sequence of interactions is denoted as  $[(x_1,r_{x_1}),\cdots,(x_T,r_{x_T})]$; each element in the sequence is an item-rating pair denoted as a tuple, where $x_t\in[M]$ is the $t^{\tth}$ item they interacted with and $r_{x_t}$ is the \emph{rating} they gave to the item $x_t$. 
The sketching policy keeps a sketch of $K$ pairs. 
Therefore, the sketch at time $t$ is denoted as $\CS_t=\{(x_{(1)},r_{x_{(1)}}),\cdots (x_{(K)},r_{x_{(K)}})\}$ where $x_{(k)} \in x_{1:t}$ and $k\in [K]$; {for the first $K$ time steps, the sketch $\CS_t$ contains all the past history.} 
Our goal is to predict their (real/binary/categorical-valued) rating on the item they interact with at the next time step, $r_{x_{t+1}}$, using the sketch  
$\CS_t$, \emph{given} that we know which item they interact with next. 
In the {\bf implicit} RS setting, for a user, the sequence of interactions is denoted as $[x_1,\cdots,x_T]$ where $x_t\in [M]$ is the item they interacted with at time $t$. 
There are no explicit ratings; items the user interacts with are considered positively rated while items that the user does not interact with are considered to be negatively rated. 
The sketching policy keeps a sketch of $K$ items. 
Therefore, the sketch at time $t$ is denoted as $\CS_t=\{x_{(1)},\cdots x_{(K)}\}$ where $x_{(k)}\in x_{1:t}$ and $k\in [K]$. 
Our goal is to predict the item that they interact with next, $x_{t+1}$, out of the entire set of items $[M]$, using the sketch $\CS_t$.

We consider two sketching policy updating settings depending on how frequently the sketch $\CS_t$ is updated.
In the {\bf online setting}, we update the sketch at each time step. 
Specifically, given the sketch at the last time step, $\CS_{t-1}$ and the current interaction/item $(x_t,r_{x_t})$ (or $x_t$) for the explicit (or implicit) case at time $t$, the sketching policy decides whether to include the current item in the sketch; if so, it also decides which item to remove from $\CS_{t-1}$ to keep the size of the sketch fixed, arriving at the sketch for the current time step, $\CS_t$.
We use $\hat{\CS}_{t} = \CS_{t-1} \cup \{(x_t,r_{x_t})\}$ (or $\{x_t\}$) to denote the \emph{intermediate} sketch of $K+1$ items and the sketching policy decides which \emph{single} item to remove from $\hat{\CS}_{t}$ to get the new sketch $\CS_t$. 
In the {\bf batch setting}, we update the sketch once every $\tau$ time steps; setting $\tau=1$ results in the online setting. 
Our method is equally applicable to the case of varying update time periods $\tau_1,\tau_2,\cdots$;
for notation simplicity, we will only detail the case of a fixed time period $\tau$ in this paper.
Specifically, given the sketch at the last time step, $\CS_{t}$, and a batch of current interactions/items $[(x_{t+1},r_{x_{t+1}}),\cdots,(x_{t+\tau},r_{x_{t+\tau}})]$ (or $[x_{t+1},\cdots,x_{t+\tau}]$) for the explicit (or implicit) case at time $t+\tau$, the sketching policy decides whether to include the current items in the sketch and in that case which items to remove, arriving at the sketch for the current time step $\CS_{t+\tau}$. 
Similarly, we use $\hat{\CS}_{t+\tau} = \CS_{t} \cup \{(x_{t+1},r_{x_{t+1}}),\cdots,(x_{t+\tau},r_{x_{t+\tau}})\}$ (or $\{x_{t+1},\cdots,x_{t+\tau}\}$) to denote the intermediate sketch of $K+\tau$ items and the sketching policy decides which $\tau$ items to remove from $\hat{\CS}_{t+\tau}$ to get the new sketch $\CS_{t+\tau}$.

\subsection{Sketching Objective}
We solve the following bilevel optimization problem (for one user only for notation simplicity) \cite{bilevel}:

\begin{align}
    &\underset{\Theta,\Phi}{\text{minimize}} \sum_{t=0}^{T-1} \ell(r_{x_{t+1}}\!,\! g(x_{t+1}\!;\!\theta_t^{\ast}(\Theta\!,\!\Phi)))\!\triangleq\!\sum_{t=0}^{T-1}\!\ell_{t+1}(\!\theta_{t}^{\ast})\!\label{eq:bilevel} \\
    &\text{s.t.}\,\theta_t^{\ast}\!\!=\!\!\argmin_{\theta_t}\!\sum_{k=1}^K\! \ell( r_{x_{(k)}}\!,\!g(x_{(k)};\theta_t))\!+\!\CR\!(\theta_t\!;\!\Theta)\!\triangleq\!\CL\!(\CS_t;\theta_t)\label{eq:inner}\\
    &\text{where}\,\CS_{t+1:t+\tau}=\pi(\CS_t,(x_{t+1:t+\tau},r_{t+1:t+\tau});\Phi). \label{eq:algo}
\end{align}
Here, $\Theta$ and $\Phi$ are the global RS model and sketching policy parameters, respectively. 
$g(\cdot)$ is the RS model that takes an item $x_{t}$ as input and predict its explicit or implicit rating (which we denote as $r_{x_t}=1$). %
$\pi(\cdot)$ is the sketching policy that takes as input the sketch at the last time step, $\CS_{t}$, the current items $x_{t+1:t+\tau}$, and outputs the updated sketch $\CS_{t+\tau}$.

The outer-level optimization problem minimizes the loss, $\ell(r_{x_{t+1}}, g(x_{t+1};\theta_t^{\ast}))$ across all users and all time steps to learn both the global RS model and the sketching policy. 
The inner-level optimization problem minimizes $\CL(\CS_t;\theta_t)$, the loss on the sketch for each user at each time step, to adapt the global RS model locally, resulting in a user, time step-specific parameter ${\theta}_t^{\ast}$. $\CR({\theta}_t;{\Theta})$ is a regularization term that penalizes large deviations of the local parameters from global values. 
Note that ${\theta}_t^*$ is a function of the global parameters ${\Theta}$ and ${\Phi}$, reflected through both the regularization term in (\ref{eq:inner}) and the items the sketching policy selects for the user in (\ref{eq:algo}).

\subsection{Recommender System Model}
Since our focus in this paper is differentiable sketching policy learning, which is agnostic to the underlying base RS model, we adopt a standard neural collaborative filtering (NCF) model as the base RS model \cite{ncf} since NCF works well with gradient-based optimization;
we use NCF to compute the loss $\ell(r_{x_t}, g(x_t;\theta_t))$ in both the inner and outer optimization problems. We emphasize that our approach is model agnostic and equally applicable to any differentiable RS model; in the experiments, we also use case studies to show that a learned sketching policy under one RS model is still highly effective for another RS model.

The prediction model parameter $\Theta$ contains the embedding of a user $\Theta(u)$ and a neural network with parameter $\Theta(p)$ corresponding to the parameters of the items. 
For simplicity, we will use $\Theta$ to denote $\{\Theta(u),\Theta(p)\}$. 
For the explicit RS setting, given the local parameter $\theta_t$ and the next item $x_t$, we predict the rating $r_t$ as $g(x_t;\theta_t)$. For real-valued ratings, we define a Gaussian likelihood function and use the mean-squared error loss $\ell_{\text{mse}}$; for binary (or categorical) ratings, we define a logistic (softmax) likelihood function resulting in the binary (or categorical) cross-entropy loss $\ell_{\text{bce}}$ (or $\ell_{\text{cce}}$).  
For the implicit RS setting, we predict the next item $x_t$ as $g(x_t; \theta)$ among all the items $[M]$.
We define a softmax function over all $M$ items, resulting in a categorical cross-entropy loss. The number of items is often large; therefore, several alternative loss functions such as the bayesian personalized ranking loss or the Top1 loss, together with negative sampling, are often used instead \cite{bpr,gru4rec}.
We emphasize that our method is agnostic to the loss function; for simplicity, we use  the standard categorical cross-entropy $\ell_{\text{cce}}$ loss in our experiments.

\subsection{Sketching Policy}
We use a sparse vector $\bz_t\in \{0,1\}^M$ to represent the indices of each item in the sketch $\CS_t$ at time $t$, with $\bz_{t,j}=1$ if and only if item index $j$ is present in the current sketch. This vector has a one-to-one correspondence with the sketch $\CS_t$.
We also use the vector $\by=[r_1,\cdots,r_M]\in \BR^M$ to represent the user's ratings of all items. 
These ratings are real-valued under the explicit RS setting and binary-valued under the implicit RS setting.\footnote{Our framework allows multiple interactions with the same item; for notation simplicity, we detail our method in the case where a user interacts with each item at most once.} The ratings on the non-interacted items do not need to be defined; the DiPS algorithm masks ratings on these items. 
In the online ($\tau=1$) and batch update settings, using the intermediate sketch we defined above,  at time $t+\tau$, we have
\begin{align*}
\hat{\bz}_{t+\tau} = \bz_{t}+ \sum_{j=t+1}^{t+\tau}\be_{x_{j}},
\end{align*}
where $\be_{x_j}\in \{0,1\}^M$ represents the unit vector with a $1$ only at index $x_j$ and $0$ at all other indices. 

The sketching policy %
$\pi$ only has access to items in the intermediate sketch. 
Therefore, we can represent this rating information using the vector $\hat{\bz}_{t+\tau}\odot \by\in \BR^M$ where $\odot$ denotes element-wise multiplication. 
The policy $\pi(\hat{\bz}_{t+\tau}, \by;\Phi)$ updates the intermediate sketch $\hat{\CS}_{t+\tau}$ to $\CS_{t+\tau}$. 
In particular, it outputs a sparse vector $\bw_{t+\tau}\in \{0,1\}^M$ that indicates whether each item in the sketch should be kept or removed. 
In the online setting, the policy outputs the item index to \emph{remove}, $\bw_{t+1}\in \{0,1\}^M\cap \Delta^{M-1}$, where $\Delta^{M-1}$ is the probability simplex. 
In the batch setting, the policy outputs the $K$ item indices to \emph{keep}, $\bw_{t+\tau}\in \{0,1\}^M\cap \{\bw: 1^T\bw=K\}$. %

The sketching policy $\pi$ computes a score for each item that is in the intermediate sketch $\hat{\bz}_{t+\tau}$ using a neural network $f(\cdot)$ with the observed ratings as $f(\hat{\bz}_{t+1}, \by;\Phi)= f(\hat{\bz}_{t+1}\odot \by;\Phi)$. 
In the online setting, 
we use the softmax distribution $\sigma(f(\hat{\bz}_{t+1}\odot\by;\Phi))$ to select the item to remove, $\bw_{t+1}(\sigma(f(\hat{\bz}_{t+1}\odot\by;\Phi)))$. 
We can do this either in a deterministic way by selecting the item with the highest score or in a stochastic way by sampling from the softmax probability distribution. 
The item indices included in the updated sketch $\CS_{t+1}$ are then computed as
\begin{align}
\bz_{t+1}  ={\bz}_{t}+\be_{x_{t+1}}- \bw_{t+1}(\sigma(f(\hat{\bz}_{t+1}\odot\by;\Phi))).\label{eq:softmax-layer} 
\end{align}
In the batch setting, we need to select $K$ items from $K+\tau$ items. 
We employ the Top-K projection layer \cite{topk-projection} defined as 
\begin{align}
  \mu(f(\hat{\bz}_{t+\tau}\odot \by;\Phi))&=\!\argmin_{0<\bu<1} -\!f(\hat{\bz}_{t+\tau}\!\odot\! \by;\Phi)^T\!\bu \!-\!H(\bu)\nonumber\\ 
  &  \mbox{s.t.}\,\,1^T\bu=K,
\end{align}
where $H(\bu)$ is the binary cross entropy function and $f(\hat{\bz}_{t+\tau}\odot \by;\Phi)$ is the score for the $M$ items.
Similarly, we can sample the $K$ points to keep, $\bw_{t+\tau}(\mu(f(\hat{\bz}_{t+\tau},y;\Phi)))$, in either a deterministic way or a stochastic way.
The sketch at time $t+\tau$ is given by
\begin{align}
\bz_{t+\tau} = \bw_{t+\tau}(\mu(f(\hat{\bz}_{t+\tau}\odot \by;\Phi))).    \label{eq:topk-layer}
\end{align}
In both cases, the sketching policy output is only defined over items in the intermediate sketch. This constraint 
can be satisfied by adding $\log \hat{\bz}_{t+\tau}$ as input to the final softmax or Top-K projection layer of the sketching policy network. 

\subsection{Optimization}
At the inner-level, we adapt the user parameter $\theta_t^{\ast}$ from the global parameter $\Theta$ using the sketched items $\CS_t$ at each time step. 
In practice, we keep item-specific neural network parameters $\Theta(p)$ fixed and adapt only the user embedding $\Theta(u)$ to minimize the loss on the $K$ items in the sketch. Following the model agnostic meta learning approach \cite{maml}, we set $\theta_t(u),\theta_t(p)\leftarrow \Theta(u), \Theta(p)$ and take a fixed number of gradient descent (GD) steps as
\begin{align}
\theta_t(u)\leftarrow \theta_t(u) -\alpha\nabla_{\theta(u)}\CL(\CS_t;\theta)|_{\theta=\theta_t}.\label{eq:gd-steps}
\end{align}
A fixed number GD steps in Eq.~\ref{eq:gd-steps} is equivalent to implicit regularization \cite{bayes-regularization}; thus, we do not impose any explicit regularization in the inner optimization problem. 
Since $\theta_t^{\ast}$ is a function of $\Theta$,
computing the gradient w.r.t.\ $\Theta$ in the outer optimization objective (\ref{eq:bilevel}) requires us to compute the gradient w.r.t.\ the gradient in (\ref{eq:gd-steps}), i.e., the meta gradient, which can be computed using automatic differentiation \cite{automatic}. 
Similarly, to learn the sketching policy parameters $\Phi$, we need to compute the gradient of the outer optimization objective w.r.t.\ $\Phi$ through the user parameters $\theta_t^{\ast}(\Theta,\Phi)$ in (\ref{eq:inner}). 
However, the discrete item indices to remove from the intermediate sketch are non-differentiable. 
Therefore, we need to develop a method to approximate this gradient, which we detail next.

\subsubsection{Sketching Policy Optimization.}
 \begin{figure}[tp]
    \centering
    \includegraphics[width=0.49\textwidth]{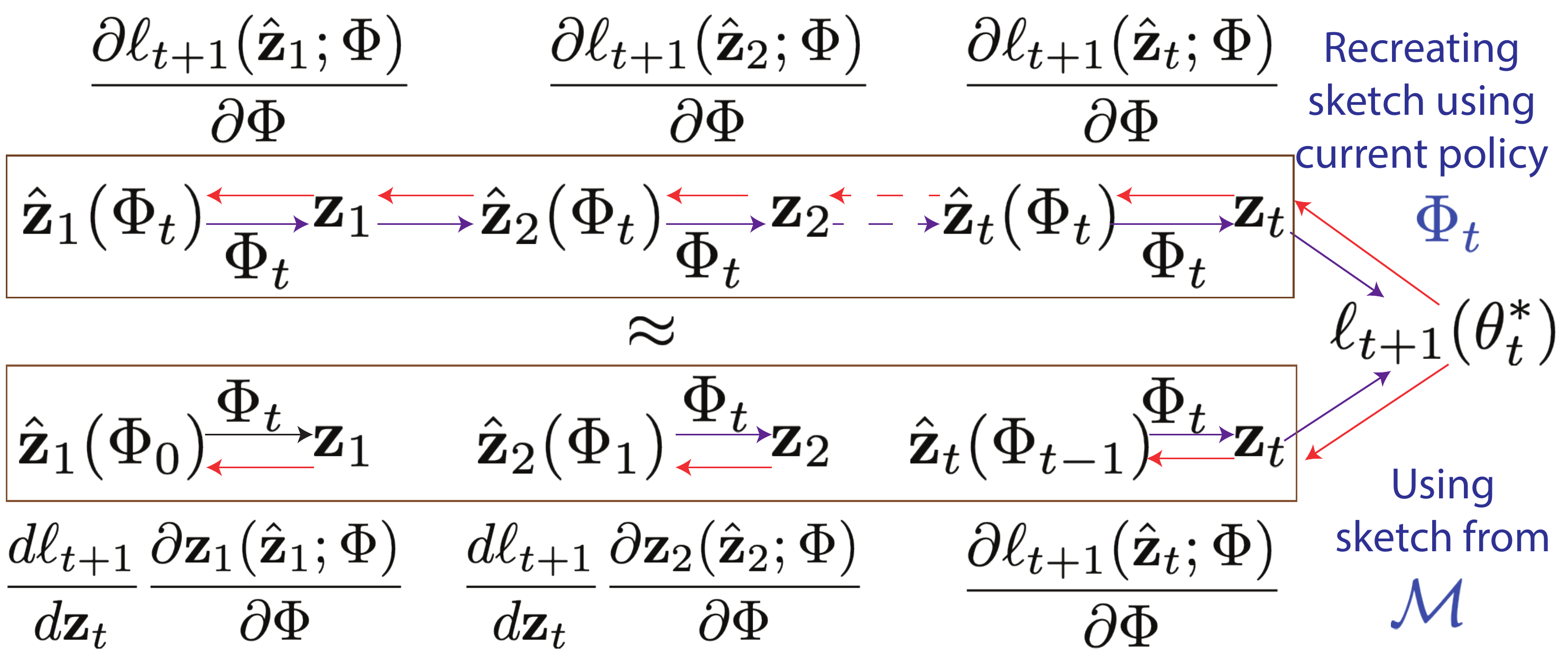}
\caption{Top/bottom: true/approximate gradient computation at time step $t$.
The approximate gradient calculated using intermediate sketches, obtained from past parameters $\Phi_{1:t-1}$, is close to the true gradient when the learning rate is small.}
    \label{fig:grad-simple}
\vspace{-0.2in}
\end{figure}
The inner-level optimization in (\ref{eq:inner}) uses ${\bz}_t$, the vector version of the sketch $\CS_t$, to compute the inner-level loss, which is used to adapt the user specific parameter $\theta_t^{\ast}$. 
This loss is computed on all items, regardless of whether they are part of the sketch, and multiplied with the weight vector $\bz_t$ before taking gradient steps. 
Therefore, we can still compute the gradient w.r.t.\ to the weight of all the items $\frac{d \ell_{t+1}}{d {\bz}_t}$ even if their corresponding weight is zero.
We start with the online setting and denote the outer optimization objective at time $t+1$ as $\ell_{t+1}(\theta_{t}^{\ast})$. 
Thus, we need to compute $\frac{d\ell_{t+1}}{d\Phi}$. Note that $\ell_{t+1}$ is a function of $\theta_t^{\ast}$, which is a function of $\Phi$ (from (\ref{eq:inner}) and (\ref{eq:algo})) as
\begin{align}
\theta_t^{\ast}=\argmin_{\theta_t}\sum_{j=1}^M \bz_{t,j}(\Phi)\ell(r_j,g(j,\theta_t))+ \CR(\theta_t;\Theta). \label{eq:inner-opt}
\end{align}
We can re-write the gradient using the chain rule as 
\[\frac{d\ell_{t+1}}{d\Phi} = \frac{d\ell_{t+1}}{d\theta_t^{\ast}}\frac{d\theta_t^{\ast}}{d\Phi}=\frac{d\ell_{t+1}}{d\theta_t^{\ast}}\frac{d\theta_t^{\ast}}{d\bz_t} \frac{d\bz_t}{d\Phi}.\]
We also note that the sketch item indices $\bz_t$ at time $t$ is a function of $\{\bz_{t-1}, \cdots, \bz_1\}$ which are themselves a function of $\Phi$.
We can write the total derivative of $\bz_t$ w.r.t.\ $\Phi$ in terms of the partial derivative as
\begin{align}
    \frac{d\bz_t}{d\Phi} \!=\!\frac{\partial\bz_t(\hat{\bz}_t\!\odot\!\by;\Phi)}{\partial\Phi} \!+\! \sum_{j=1}^{t-1}\! \frac{\partial\bz_j(\hat{\bz}_j\!\odot\!\by;\Phi)}{\partial\Phi}\!(\!\prod_{l=j}^{t-1}\! \frac{d\bz_{l+1}}{d\bz_{l}}\!),\label{eq:full-grad}
\end{align}
where the partial derivatives $\frac{\partial \bz_j(\hat{\bz}_j;\Phi)}{\partial \Phi}$ w.r.t.\ $\Phi$ are computed by keeping the input $\hat{\bz}_j$ constant.
The main challenge here is that in order to compute the gradient for the loss on item $x_{t+1}$, we need to re-generate the computation graph from $\bz_1$ to $\bz_t$, i.e., the entire sketching history, using the \emph{current} policy parameter $\Phi$ (at time step $t$), which cannot be computed in previous time steps with past policy parameters (with multiple SGD steps in between). 
This regeneration is often infeasible due to its high computational cost. An alternative is to run the sketching with the current policy parameter and solve the inner optimization for each time step  at once; however, that leads to enormous memory requirements for the backward gradient propagation even for a few time steps.

We propose to approximate the total derivative in (\ref{eq:full-grad}) without recomputing the entire sketching process at every time step. For each iteration, we take a mini-batch of users (with multiple time steps for multiple interactions) for stochastic gradient descent (SGD) optimization. We represent the policy parameter $\Phi$ at time $t$ as $\Phi_t$ in the training iteration.
Note that in (\ref{eq:full-grad}), every past $\bz_j$ (and $\hat{\bz}_j$) correspond to the sketch indices obtained using the current parameter $\Phi_t$. 
However, we can use a queue $\CM$ storing the intermediate sketch indices $\CM=[\hat{\bz}_1,\cdots, \hat{\bz}_{t}]$ computed from old policy parameters, $\Phi_0, \Phi_1,\cdots,\Phi_{t-1}$ respectively.
If the learning rate is small enough in the SGD steps, we can assume that the past sketches stored in queue, which were computed from the old sketching policy parameters, to be close to that computed from the new parameters $\Phi_t$. 
At time step $t$, we can then run the sketching policy with the current parameter $\Phi_t$ on the stored intermediate sketch indices $\CM$ in parallel to obtain $\bz_{1:{t-1}}$ and  compute $\frac{\partial\bz_j(\hat{\bz}_j\odot \by;\Phi)}{\partial\Phi}|_{\Phi=\Phi_t}$ efficiently. 
We can approximate the Jacobian $\frac{d\bz_{l+1}}{d\bz_l}$ with the identity matrix since they are additive in (\ref{eq:softmax-layer}), which does not need to be explicitly generated in (\ref{eq:full-grad}). 
We can compute the gradient $\bv=\frac{d\ell_{t+1}}{d\bz_t}$ and obtain the vector-Jacobian product $\frac{d\ell_{t+1}}{d\bz_t}\frac{d\bz_t}{d\Phi}$ efficiently as
\begin{align}
 \frac{d\ell_{t+1}}{d\Phi} \approx \frac{\partial\ell_{t+1}(\hat{\bz}_t;\Phi)}{\partial\Phi} +\frac{\partial}{\partial \Phi}  \Big(\bv^T\sum_{j=1}^{t-1}(\bz_j(\hat{\bz}_j;\Phi)\Big),\label{eq:approx-grad}
\end{align}
where $\bv$ is fixed and only $\bz_j$'s are a function of $\Phi$ for computing the partial derivatives. We note that in (\ref{eq:approx-grad}), there is no sequential dependency; all the terms can be computed in parallel.
We use a fixed-size queue $\CM$ (with size $Q\sim 50-100$) where we remove the oldest sketch indices when the queue gets full. 
We update the queue after every $\tau$ time steps. This approximate gradient computation process is visualized in Figure~\ref{fig:grad-simple}. In the supplementary material, we show that this approximate gradient remains close to the true gradient.

Since the sketch item indices $\bz_t(\hat{\bz}_t,\by,\Phi)$ are sampled from the Softmax or Top-K projection layer, they are not differentiable w.r.t the policy parameters $\Phi$. Thus, we need to approximate the partial derivative $\frac{\partial\bz_t(\hat{\bz}_t,\by,\Phi)}{\partial\Phi}$,
which  we can re-write in the online setting as
\[\frac{\partial\bz_t(\hat{\bz}_t,\by;\Phi)}{\partial\Phi} =\frac{d\bz_t}{d\bw_t}\frac{d\bw_t}{d\sigma(f(\cdot))} \frac{\partial\sigma(f(\hat{\bz}_{t}\odot\by;\Phi))}{\partial\Phi}, \]
where $\sigma$ is the softmax layer and $\bw_t$ contains the sketch indices after sampling in (\ref{eq:softmax-layer}).
We need to approximate $\frac{d\bw_t}{\sigma(f(\cdot))}$ since they are non-differentable; we can leverage the approximation $\bw_t\approx\sigma(f(\cdot))$ since it holds if the item to be removed has almost all the probability mass. 
This approximation is known as the straight-through (ST) estimator and it is often found to have lower empirical variance than the REINFORCE gradient estimator \cite{reinforce,st}. 
In general,  one can test
other differentiable approximations, such as ST-Gumbel softmax estimator \cite{gumbel}, for the sampling operation in (\ref{eq:softmax-layer}) and (\ref{eq:topk-layer}), which we leave for future work.
The final term
$\frac{\partial\sigma(f(\hat{\bz}_{t}\odot\by;\Phi))}{d\Phi}$ can be easily computed as the gradient of the softmax layer w.r.t.\ the policy parameters. 
In the batch setting, 
the softmax layer ($\sigma$) is replaced by the Top-K projection layer ($\mu$) that scores top $K$ items close to $1$ and other items close to $0$. 
We can approximate the second term as $\bw_t\approx\mu(f(\cdot))$ when the top $K$ items that are selected have the highest scores among all items.  %
We can further use the KKT conditions and the implicit function theorem to compute the gradient $\frac{\partial\mu(f(\hat{\bz}_{t}\odot\by;\Phi))}{d\Phi}$; for details, refer to  \cite{meta-opt-fix,topk-projection}.

\paragraph{Connection to Influence Function.}
In (\ref{eq:inner-opt}), we can compute the gradient of $\theta_t^{\ast}$ w.r.t.\  $\bz_{t,j}$, $\forall j\in x_{1:t}$ using the implicit function theorem \cite{cook1982residuals} as
\begin{align}
    \frac{d\theta_t^{\ast}}{d\bz_{t,j}} \!=\!-(\nabla^2_{\theta_t}\CL(\CS_t;\theta_t))^{-1}\nabla_{\theta_t}\ell(r_j,g(j,\theta_t))|_{\theta_t=\theta_t^{\ast}}.
\end{align}
The gradient for loss on the next rating prediction is given by
\begin{align*}
\frac{d\ell_{t+1}}{d\bz_{t,j}(\Phi)}=-(\nabla_{\theta_t}\ell(r_{x_{t+1}};g(x_{t+1};\theta_t)))(\nabla^2_{\theta_t}\CL(\CS_t;\theta_t))^{-1}\\\nabla_{\theta_t}\ell(r_j,g(j,\theta_t))|_{\theta_t=\theta_t^{\ast}}:=\CI_{t+1}(j),
\end{align*}
where $\CI_{t+1}(j)$, the influence function \cite{influence} score of item $j$, computes the change in the loss on the next time step under
small perturbations in the weight of this item, $\bz_{t,j}$ in (\ref{eq:inner-opt}). Intuitively, we would want to keep items that have gradients similar to that for the future items in sketch, i.e., those that are the most informative of future recommendations. 
Therefore, in the online setting, the sketching policy will tend to select items that are the least informative (to replace from the sketch) and in the batch setting, it will tend to select items that are most informative (to keep in the sketch).

\begin{algorithm}[t]
\begin{algorithmic}[1]
\STATE Initialize global parameters ${\Theta}, {\Phi}$, learning rates $\eta$ (outer level), $\alpha$ (inner level) sketch size $K$, queue size $Q$.%
\WHILE {not converged} 
\STATE Randomly sample a mini-batch of  $n$ users.
\STATE For each user, initialize empty queue of past sketch indices $\CM\leftarrow \phi$, sketch $\CS_0$ and sketch indices $\bz_0\in \{0,1\}^{M}$, encode ratings into vector $\by \in \BR^{M}$.
\FOR{$t\in 1\cdots (T-1)$} 
\STATE  For each user, optimize $\theta_t^{\ast}$ on the sketch $\CS_{t-1}$.
\STATE Compute loss $\ell_{t+1}$ on item $(x_{t+1}, r_{x_{t+1}})$ using $\theta_t^{\ast}$.
\STATE Compute $\nabla_{\Theta}\ell_{t+1}$ update ${\Theta}$:  ${\Theta}\!\leftarrow\! {\Theta}\!-\!\eta\nabla_{{\Theta}}\ell_{t+1}$.
\STATE Computed intermediate indices $\hat{\bz}_t\leftarrow{\bz}_{t-1}+\be_{x_t}$.
\IF{$t>K$}
\STATE Compute $\bz_j$ for $j\in\{1,\cdots,t-1\}$ in parallel from stored $\hat{\bz}_j$ in queue $\CM$ using $\pi(\cdot;\Phi)$.
\STATE Compute $\nabla_{\Phi}\ell_{t+1}$ using Eq.\ref{eq:approx-grad} and update ${\Phi}$:  ${\Phi}\leftarrow {\Phi}-\eta\nabla_{{\Phi}}\ell_{t+1}$.
\STATE Append $\hat{\bz}_t$ into queue $\CM$, remove oldest if full.
\ENDIF
\STATE Compute $\CS_t$ and $\bz_t$ using policy $\pi$ or set $\bz_t\leftarrow\hat{\bz}_t$.
\ENDFOR
\ENDWHILE 
\end{algorithmic}
\caption{Training of DiPS}
\label{alg:training}
\end{algorithm}

\section{Experimental Results}
 \begin{table}[t]\centering
 \scalebox{0.7}{
     \begin{tabular}{llllll}\toprule
     Dataset & Movielens 1M & Movielens 10M & Netflix & Book & Forusquare\\\midrule
    Users & 6K & 70K & 430K & 22K  & 52K \\
    Items & 3.7K & 11K & 18K &24K & 37K \\
    Interactions &1M & 10M & 100M & 1.1M & 2.3M \\\midrule
     \end{tabular}
}
\caption{Dataset Statistics
}
\label{tab:dataset}
 \end{table}

\paragraph{Datasets and Evaluation Metric.}
We use five publicly available benchmark datasets: the 
Movielens 1M\footnote{\url{https://grouplens.org/datasets/movielens/1m/}} and 10M \footnote{\url{https://grouplens.org/datasets/movielens/10m/}} datasets   \cite{movielens} and the Netflix Prize dataset \footnote{\url{https://www.kaggle.com/netflix-inc/netflix-prize-data},\url{https://www.netflixprize.com/}} for explicit RSs and the Amazon Book\footnote{\url{https://jmcauley.ucsd.edu/data/amazon/}} and Foursquare\footnote{\url{https://sites.google.com/site/yangdingqi/home/foursquare-dataset}} datasets for implicit RSs. The Movielens datasets contain at least 20 ratings for each user; for the Netflix dataset, we filter out users with less than 20 ratings.  
We use 20-core settings for the Foursquare and the Amazon book dataset; thus, all users (items) interact with at least 20 items (users). For Amazon Book dataset, we keep reviews with ratings more than 3.5 (from 1-5 scale) as the implicit positively rated items \cite{book}. The foursquare dataset contains global user check-in datasets on the Foursquare platform from Apr.\ 2012 to Jan.\ 2014 \cite{foursquare}. See Table~\ref{tab:dataset} for detailed statistics. 
For explicit RSs, we use root mean square error (RMSE) as the evaluation metric. For implicit RSs, we use Recall@20 ($=\BE \mathbbm{1}_{\text{rank}\leq K}$) as the evaluation metric where rank is computed among all possible items; we also provide additional results with Mean Reciprocal Rank (MRR)@20 as the evaluation metric where MRR@K = $\BE \frac{ \mathbbm{1}_{\text{rank}\leq K}}{\text{rank}}$. 
We randomly split $60$-$20$-$20\%$ of the users in the datasets into training-validation-testing sets. We run all experiments five times with different splits and report the average and standard deviation (std) numbers across all five runs.

\paragraph{Methods and Baselines.}
 \begin{table*}[t]\centering
 \scalebox{0.9}{
     \begin{tabular}{ccc|cccccc}\toprule
     Settings ($\tau$) & Dataset & K & Random & Hardest & Influence & DiPS@1 & DiPS\\\midrule

\multirow{9}{*}{Online (1)} &\multirow{3}{*}{Movielens 1M} & 2& 0.9701$\pm$0.0015& 0.9747$\pm$0.0026& 0.9747$\pm$0.0016& 0.9615$\pm$0.002& \bf{0.9543$\pm$ 0.0015}\\
 && 4& 0.955$\pm$0.0023& 0.9606$\pm$0.0025& 0.9718$\pm$0.0008& 0.949$\pm$0.0014& \bf{0.9418$\pm$ 0.0018}\\
 && 8& 0.9387$\pm$0.0017& 0.9435$\pm$0.0024& 0.9662$\pm$0.0016& 0.9354$\pm$0.0017& \bf{0.93$\pm$ 0.002}\\
\cline{2-8}
&\multirow{3}{*}{Movielens 10M} & 2& 0.9232$\pm$0.0011& 0.9221$\pm$0.0013& 0.9147$\pm$0.0015& 0.9174$\pm$0.001& \bf{0.9008$\pm$ 0.0011}\\
 && 4& 0.9065$\pm$0.0011& 0.903$\pm$0.0014& 0.9054$\pm$0.0017& 0.9009$\pm$0.0011& \bf{0.8874$\pm$ 0.0011}\\
 && 8& 0.8853$\pm$0.001& 0.8808$\pm$0.0013& 0.8948$\pm$0.0016& 0.8812$\pm$0.0011& \bf{0.8726$\pm$ 0.0008}\\
 \cline{2-8}
 & \multirow{3}{*}{Netflix} & 2& 0.9898$\pm$0.0006& 0.9946$\pm$0.0008& 0.9826$\pm$0.001& 0.9807$\pm$0.0009& \bf{0.9646$\pm$ 0.0005}\\
 && 4& 0.9708$\pm$0.0006& 0.9786$\pm$0.0008& 0.9726$\pm$0.0012& 0.9631$\pm$0.0009& \bf{0.9532$\pm$ 0.0006}\\
 && 8& 0.9474$\pm$0.0006& 0.9564$\pm$0.0007& 0.9598$\pm$0.0013& 0.9411$\pm$0.0007& \bf{0.9351$\pm$ 0.0006}\\
\midrule
\midrule
\multirow{9}{*}{Batch (4)} &\multirow{3}{*}{Movielens 1M} & 2& 0.9741$\pm$0.0022& 0.9965$\pm$0.0024& 0.997$\pm$0.0016& 0.9658$\pm$0.0023& \bf{0.9651$\pm$ 0.0027}\\
 && 4& 0.9611$\pm$0.0025& 0.9848$\pm$0.0021& 0.9867$\pm$0.0016& \bf{0.9537$\pm$ 0.0027}& 0.9592$\pm$0.0017\\
 && 8& 0.9455$\pm$0.0017& 0.9678$\pm$0.0017& 0.9753$\pm$0.0017& \bf{0.9418$\pm$ 0.002}& 0.9476$\pm$0.0018\\
\cline{2-8}
&\multirow{3}{*}{Movielens 10M} & 2& 0.9221$\pm$0.0012& 0.9271$\pm$0.0015& 0.9127$\pm$0.0013& 0.9145$\pm$0.0011& \bf{0.9086$\pm$ 0.0011}\\
 && 4& 0.9055$\pm$0.0014& 0.9098$\pm$0.0013& 0.9016$\pm$0.0014& 0.9001$\pm$0.0013& \bf{0.8788$\pm$ 0.001}\\
 && 8& 0.8839$\pm$0.0014& 0.888$\pm$0.0014& 0.8907$\pm$0.0015& 0.8801$\pm$0.0015& \bf{0.8647$\pm$ 0.001}\\
\cline{2-8}
&\multirow{3}{*}{Netflix} & 2& 0.9825$\pm$0.0006& 0.9925$\pm$0.0008& 0.9715$\pm$0.0007& 0.9732$\pm$0.0008& \bf{0.9632$\pm$ 0.0007}\\
 && 4& 0.9625$\pm$0.0005& 0.9766$\pm$0.0007& 0.9618$\pm$0.0007& 0.955$\pm$0.0005& \bf{0.919$\pm$ 0.0008}\\
 && 8& 0.9387$\pm$0.0005& 0.9543$\pm$0.0007& 0.9497$\pm$0.0007& 0.9327$\pm$0.0005& \bf{0.909$\pm$ 0.0005}\\
\midrule
     \end{tabular}
}
\caption{Mean and std RMSE for all methods under the online ($\tau=1$) and batch setting ($\tau=4$) on all explicit RS datasets.
}
\label{tab:rmse}
 \end{table*}

   \begin{table*}[t]\centering
 \scalebox{0.9}{
     \begin{tabular}{ccc|cccccc}\toprule
     Settings ($\tau$) & Dataset & K & Random & Hardest & Influence & DiPS@1 & DiPS\\\midrule

\multirow{6}{*}{Online ($1$)}&\multirow{3}{*}{Book} & 2& 0.0672$\pm$0.0002& 0.0682$\pm$0.0004& 0.074$\pm$0.0005& \bf{0.1244$\pm$ 0.0017}& 0.1163$\pm$0.0006\\
 & & 4& 0.0769$\pm$0.0006& 0.0787$\pm$0.0007& 0.0828$\pm$0.0004& \bf{0.1349$\pm$ 0.0005}& 0.1262$\pm$0.0007\\
 & & 8& 0.0845$\pm$0.0005& 0.0876$\pm$0.0008& 0.0877$\pm$0.0005& \bf{0.1325$\pm$ 0.0007}& 0.1275$\pm$0.0007\\
\cline{2-8}
& \multirow{3}{*}{Foursquare} & 2& 0.1329$\pm$0.0002& 0.1294$\pm$0.0003& 0.1316$\pm$0.0003& 0.1396$\pm$0.0001& \bf{0.1406$\pm$ 0.0003}\\
 && 4& 0.1416$\pm$0.0002& 0.1368$\pm$0.0003& 0.141$\pm$0.0003& 0.1512$\pm$0.0001& \bf{0.1513$\pm$ 0.0002}\\
 & & 8& 0.1508$\pm$0.0001& 0.1456$\pm$0.0002& 0.1489$\pm$0.0003& 0.1591$\pm$0.0002& \bf{0.1601$\pm$ 0.0001}\\
\midrule
\midrule
\multirow{6}{*}{Batch ($4$)}&\multirow{3}{*}{Book} & 2& 0.0564$\pm$0.0004& 0.0404$\pm$0.0003& 0.068$\pm$0.0004& \bf{0.0859$\pm$ 0.001}& 0.0751$\pm$0.0004\\
 && 4& 0.0647$\pm$0.0006& 0.0514$\pm$0.0004& 0.0787$\pm$0.0004& \bf{0.1048$\pm$ 0.0008}& 0.0987$\pm$0.0008\\
 && 8& 0.0714$\pm$0.0005& 0.0631$\pm$0.0006& 0.0844$\pm$0.0006& \bf{0.1046$\pm$ 0.0006}& 0.1011$\pm$0.0006\\
\cline{2-8}
&\multirow{3}{*}{Foursquare} & 2& 0.1203$\pm$0.0001& 0.1059$\pm$0.0003& 0.1188$\pm$0.0002& 0.1204$\pm$0.0003& \bf{0.121$\pm$ 0.0002}\\
& & 4& 0.1283$\pm$0.0001& 0.1147$\pm$0.0001& 0.1277$\pm$0.0003& \bf{0.1344$\pm$ 0.0001}& 0.1339$\pm$0.0001\\
& & 8& 0.1362$\pm$0.0002& 0.1254$\pm$0.0002& 0.1355$\pm$0.0002& 0.1406$\pm$0.0002& \bf{0.1421$\pm$ 0.0002}\\
\midrule
     \end{tabular}
}
\caption{Mean and std Recall@20 for all methods under the online ($\tau=1$) and batch setting ($\tau=4$) on all implicit RS datasets.
}
\label{tab:recall}
 \end{table*}

     \begin{table*}[t]\centering
 \scalebox{0.9}{
     \begin{tabular}{ccc|cccccc}\toprule
     Settings ($\tau$)&Dataset & K &  Random & Hardest & Influence & DiPS@1 & DiPS\\\midrule
\multirow{6}{*}{Online (1) } & \multirow{3}{*}{Book}&  2&  0.0165$\pm$0.0001& 0.0171$\pm$0.0001& 0.0201$\pm$0.0001& \bf{0.0377$\pm$ 0.0008}& 0.0331$\pm$0.0002\\
 & & 4&  0.0186$\pm$0.0002& 0.0196$\pm$0.0002& 0.0214$\pm$0.0002& \bf{0.0388$\pm$ 0.0003}& 0.0344$\pm$0.0003\\
 &  &8& 0.0209$\pm$0.0001& 0.022$\pm$0.0002& 0.0227$\pm$0.0002& \bf{0.0363$\pm$ 0.0003}& 0.0337$\pm$0.0003\\
\cline{2-8}
& \multirow{3}{*}{Foursquare} &2& 0.0342$\pm$0.0001& 0.0332$\pm$0.0001& 0.0333$\pm$0.0002& 0.0365$\pm$0.0001& \bf{0.0373$\pm$ 0.0001}\\
 && 4& 0.0367$\pm$0.0& 0.0352$\pm$0.0001& 0.0363$\pm$0.0001& 0.0403$\pm$0.0001& \bf{0.0407$\pm$ 0.0001}\\
 & & 8& 0.0397$\pm$0.0001& 0.0376$\pm$0.0001& 0.0392$\pm$0.0001& 0.0435$\pm$0.0& \bf{0.0439$\pm$ 0.0001}\\
\midrule
\midrule
\multirow{6}{*}{Batch (4) }& \multirow{3}{*}{Book} &2 &0.0128$\pm$0.0001& 0.0094$\pm$0.0001& 0.017$\pm$0.0001& \bf{0.0217$\pm$ 0.0003}& 0.0185$\pm$0.0001\\
 && 4&  0.0148$\pm$0.0001& 0.012$\pm$0.0001& 0.0194$\pm$0.0002& \bf{0.0263$\pm$ 0.0003}& 0.0244$\pm$0.0002\\
 && 8&  0.0164$\pm$0.0002& 0.0149$\pm$0.0001& 0.0206$\pm$0.0002& \bf{0.0263$\pm$ 0.0002}& 0.0247$\pm$0.0002\\
\cline{2-8}
&\multirow{3}{*}{Foursquare} &  2& 0.0309$\pm$0.0001& 0.027$\pm$0.0001& 0.0304$\pm$0.0001& 0.0315$\pm$0.0& \bf{0.0317$\pm$ 0.0001}\\
 && 4& 0.0333$\pm$0.0001& 0.0295$\pm$0.0& 0.0327$\pm$0.0& \bf{0.0362$\pm$ 0.0001}& \bf{0.0362$\pm$ 0.0001}\\
 && 8& 0.0358$\pm$0.0001& 0.0324$\pm$0.0001& 0.0351$\pm$0.0001& 0.0383$\pm$0.0001& \bf{0.0388$\pm$ 0.0001}\\
\midrule
     \end{tabular}
}
\caption{Mean and std MRR@20 for all methods under the online ($\tau=1$) and batch setting ($\tau=4$) on all implicit RS datasets.
}
\label{tab:reciprocal}
 \end{table*}

We compare our method, {\bf DiPS}, against various baselines including reservoir sampling \cite{reservoir} that has been primarily  used in RS applications, which keeps items with uniform probability \cite{streaming-reservoir,streaming-hardest}. 
We dub this heuristic sketching policy as {\bf Random}. 
There are several other heuristic sketching policies in streaming settings used in various applications. 
The {\bf Hardest} sample heuristic keeps the hardest data point to classify in the sketch and has been highly successful in continual learning and active learning tasks \cite{hard-sample}. For binary classification, it is equivalent to uncertainty sampling. 
Another closely related method is to construct a \emph{coreset} in online and batch RS settings. 
We use an {\bf Influence} function-based score to construct the sketch by selecting the most representative $K$ data points from the $K+\tau$ intermediate sketch items \cite{coresets-bilevel}. 
In contrast, our bi-level optimization framework explicitly minimizes the loss incurred on predicting \emph{future items}. 
We also experiment with a simpler version of our method, dubbed as {\bf DiPS@1}, where we do not keep the queue of past intermediate sketches and flow gradient only for the current items, i.e., the first term on the right-hand side of (\ref{eq:approx-grad}). 
We test different sketch sizes as $K \in \{2,4,8\}$. 
For the batch setting, we set the sketch update period as $\tau=4$ to cover three cases: the update period is less than, equal to, or larger than the sketch size.
Model details and parameter settings can be found in the supplementary material.

\subsection{Results and Discussion}

    \begin{figure}[tp]
    \centering
    \includegraphics[width=0.47\textwidth]{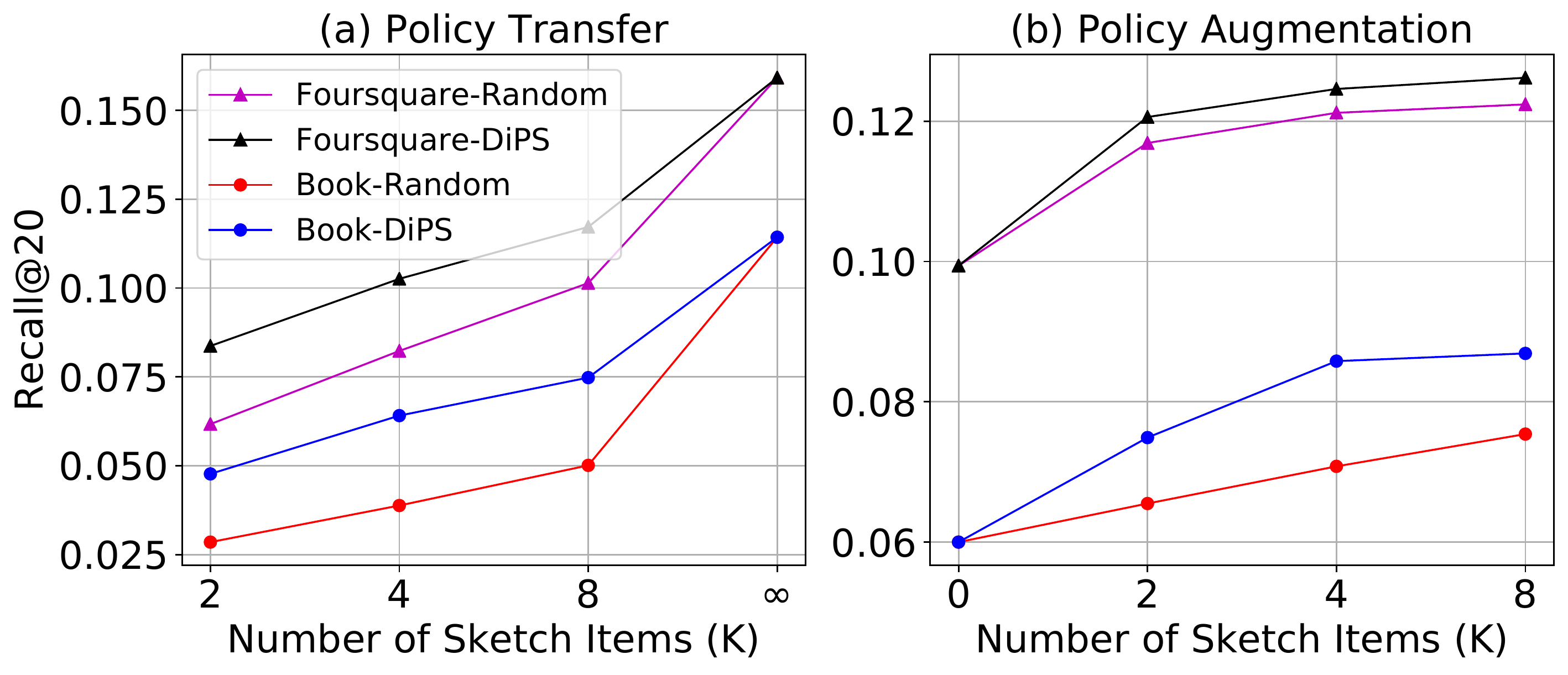}
\caption{On implicit RS datasets, (a) Recall@20 with a GRU4Rec RS model using sketching policies learned by DiPS (with NCF as the base RS model) and (b) Recall@20 with a session-based GRU4Rec model augmented with $K$ historical sketch items. %
}
    \label{fig:policy}
\vspace{-0.2in}
\end{figure}

In Table~\ref{tab:rmse}, we list the mean RMSE and std numbers across all runs for all methods on all explicit RS datasets under the online setting $(\tau=1)$.
On all datasets, for all values of the sketch size $K$, DiPS significantly outperforms other methods, followed by DiPS@1.
On all datasets, DiPS reaches similar predictive quality to that of static informativeness-based policies using up to $50\%$ fewer sketch items. 
DiPS@1 does not perform as well as DiPS, which suggests that storing past sketch steps in queue $\CM$ for more than one time step is beneficial to obtaining a more accurate gradient approximation and better predictive quality. 
We also observe that reservoir sampling slightly outperforms the Hardest and Influence heuristics on the Movielens1M and Netflix datasets while the Hardest heuristic slightly outperforms the other two on the Movielens10M dataset. 
Somewhat surprisingly, reservoir sampling performs well in many cases, without using any informativeness measures. 
We postulate that the reason behind this observation is that the Hardest and Influence heuristics operate \emph{locally} since they make decisions only in the context of the local sketch; this restriction means that they favor items that are more informative to the most \emph{recent} user interactions over those that are representative of longer-term history. 
Re-weighting these heuristics based on time difference can be beneficial and is left for future work.

In Table~\ref{tab:rmse}, we list the mean RMSE across all runs for all methods on all explicit RS datasets under the batch setting $(\tau=4)$. On all datasets, DiPS and DiPS@1 significantly outperform other informativeness-based policies. 
On the smaller datasets (Movielens 1M), DiPS@1 slightly outperforms DiPS for larger sketch sizes $K\in \{4,8\}$. 
We postulate that the reason behind this observation is that the policy is not frequently updated (only once every $\tau\!=\!4$ time steps), which reduces the benefit of more accurate gradient approximation by keeping the past sketches. Moreover, the fact that Movielens 1M is significantly smaller than the other two datasets might also contribute to this observation.

In Table~\ref{tab:recall}, we list the mean Recall@20 metric across all runs for all the methods on all implicit RS datasets under the online setting $(\tau\!=\!1)$. 
On all datasets, DiPS and DiPS@1 significantly outperform other informativeness-based policies. 
On the Book dataset, DiPS with the smallest sketch size of $K\!=\!2$ outperforms static informativeness-based policies with the largest sketch size $K\!=\!8$ by at least $30\%$. 
On the foursquare dataset, DiPS reaches similar predictive quality to that of static policies using up to $50\%$ fewer sketch items. 
We observe that DiPS@1 slightly outperforms DiPS on the smaller Book dataset. 
Combined with a similar observation in the explicit RS case, this observation suggests that storing past sketches is more beneficial on the larger datasets. 
We also observe that the Influence policy works better than other static policies on implicit RS datasets.
This observation suggests that recent context might be more important under the implicit RS setting.  
In Table~\ref{tab:recall}, we list the mean Recall@20 metric across all runs for all the methods on all implicit datasets under the batch setting $(\tau=4)$. 
On all datasets, DiPS significantly outperforms other informativeness-based policies while DiPS@1 slightly outperforms DiPS on the smaller Book dataset. 
These observations fall in line with those in the online setting.
In Table ~\ref{tab:reciprocal}, we list the mean and standard deviation of MRR@20 scores for all methods on all implicit datasets under the online setting ($\tau=1$) and batch setting ($\tau=4$).
We observe similar trends for the MRR@20 metric as the Recall@20 metric.

\paragraph{Policy Transfer.}
We perform additional experiments to show that the sketching policy learned using one base RS model, NCF in our case, can be effective even if used in conjunction with a different base RS model. 
In particular, we train a sequential GRU4Rec model \cite{gru4rec} on the Book and foursquare datasets, where at each time step $t$, all history $\bx_{1:t-1}$ is used to recommend the next item $\bx_t$.
We also train three DiPS models ($K\in \{2,4,8\}$) with a base NCF RS model (on the same training set of users) and only retain the learned sketching policies $\pi$. 
We test how this GRU4Rec model would perform when we keep a sketch of \emph{only} $K$ items $\bx_{(1):(K)}$ on the test users to recommend the next item $\bx_t$. 
In Figure~\ref{fig:policy}(a), we plot the performance of the GRU4Rec model under different values of $K$ for both the DiPS policy and the reservoir sampling policy. 
These policies are identical at $K=\infty$ when the entire history is available.
We see that the DiPS policy requires up to $50\%$ less sketch items to reach the same recommendation quality than reservoir sampling. 
This observation suggests that sketching policies learned with a particular base RS model can potentially be transferred to other base RS models effectively. 
We note that the DiPS sketching policy exploits items that are highly predictive of future items, making them amenable to other base RS models.

\paragraph{Augmented Session-based SRS.}
We perform additional experiments to show that a few historical sketch items can effectively augment base RS models to improve session-based SRS. 
In particular, we split the user's history into non-overlapping sessions of four items. At each step, the model has access to items from the current session (0-3 items) and a sketch of $K\in \{2,4,8\}$ items from the full history. 
We train a modified GRU4Rec model that computes hidden states using items from the current session plus the sketch, and concatenate the two hidden states for the final prediction layer. 
We use uniform reservoir sampling and the DiPS sketching policy (trained with NCF) to build the sketch; we train a modified GRU4Rec model on the session items and the sketched items to recommend the next item. 
In Figure~\ref{fig:policy}(b), we plot the performance of the modified GRU4Rec model. 
Note that $K=0$ represents the standard GRU4Rec model using only the session data. 
We see that augmenting historical sketch items improve the performance of the session-based GRU4Rec model by more than $20\%$ on both datasets. 
Moreover, the DiPS sketching policy achieves the same predictive quality as uniform reservoir sampling with $50\%$ fewer sketch items.

 \paragraph{Policy Visualization.} 
  \begin{figure}[tp]
    \centering
 \includegraphics[width=.5\textwidth]{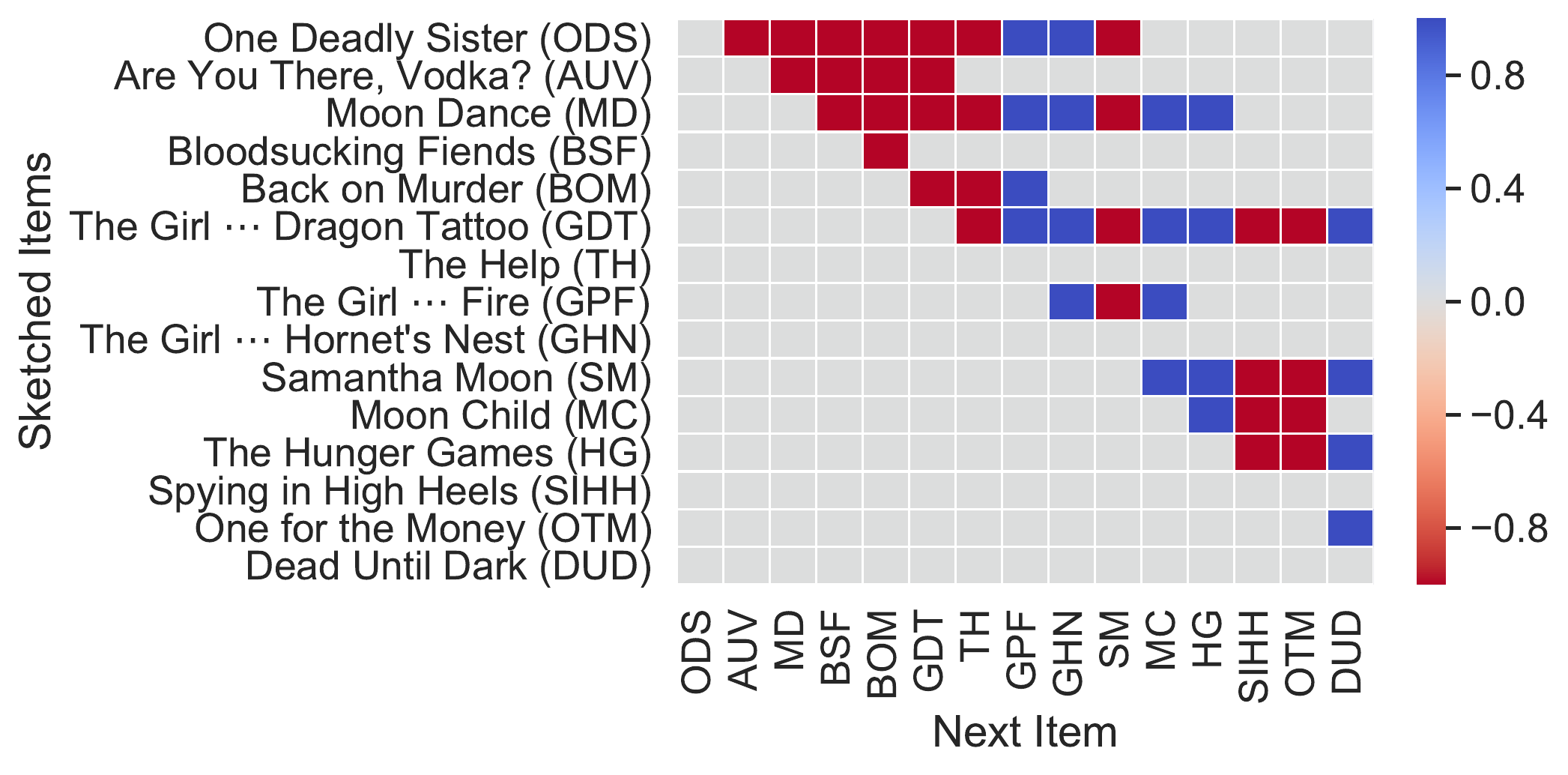}
\caption{Visualization of the DiPS sketching policy ($K=4$) on the Book dataset for a selected user over $15$ time steps. Cell $(i,j)$ represents whether item $i$ is present in the sketch and used to successfully recommend item $j$. %
}
\label{fig:policy-example}
\vspace{-0.2in}
\end{figure}
In Figure~\ref{fig:policy-example}, we plot the sketching process ($K=4$) following the policy learned by DiPS for a selected user in the Book dataset for $15$ time steps. We color-coded the columns (time steps) based on successful (blue)/unsuccessful (red) recommendations (from a total of $23,774$ distinct books), i.e., whether the actual item is included in the top-20 recommendations. 
This user is interested in the ``Mystery/Suspense'' and ``Fiction'' genres. 
The third book, ``Moon Dance'', is kept in the sketch between time steps $4$ and $12$ and used to successfully predict similar items, such as ``Moon Child''. 
The book ``The Girl with the Dragon Tattoo'' is kept in the sketching memory and used to successfully predict similar items ``The Girl Who Played with Fire'` and ``The Girl Who Kicked the Hornet's Nest''. 
The book ``The Girl Who Kicked the Hornet's Nest'' is not kept in the sketch, possibly since it is the last book in the original ``Millennium'' series; its information is already well-captured in the sketch by the first book. 
Although we cannot successfully predict the book ``Samantha Moon: The First Four Vampire'', it is kept in the sketch to capture the user's interest on fiction, which is later used to successfully predict ``Dead Until Dark'' with the same theme.
We note that although the model is not able to always successfully recommend items, the sketching policy captures item properties and adds/removes incoming items to improve future recommendations.

\section{Conclusions and Future Work}
In this paper, we developed a framework for differentiable sketching policy learning for recommender systems applications. 
The policy decides which past items to keep in a small sketch to explicitly maximize future predictive quality using items in the sketch. 
We use a bi-level optimization setup to directly learn such a sketching policy in a data-driven manner. 
Extensive experimental results on real-world datasets under various recommender systems settings show that our framework can sometimes significantly outperform existing static, informativeness-based sketching policies. 
Although side information (or metadata) often plays an important role in recommender systems, we did not use any side information in this paper. We briefly discuss how to incorporate metadata in the DiPS framework in the supplementary material.
Avenues for future work include i) using more sophisticated recommender systems models that take item metadata into account, and ii) using interpretable recommender systems architecture to explicitly interpret how past items in the sketch help us predict future items \cite{yongfeng}. 

\section*{Acknowledgements}
A.~Ghosh and A.~Lan are partially supported by the National Science Foundation via grants IIS-1917713 and IIS-2118706.

\bibliography{sketch,aaai_ref}
\newpage
\section{Supplementary Material}
\subsection{Augmenting Side Information}
In this paper, we use only past ratings $\by$ as input to the policy network and the prediction network. We note that we can also include any other side information (e.g., categories of movie and time step of when each item in the sketch was added) as input and use any differentiable neural network structure accordingly. 
For example, suppose we have item category information $\bp^1\in \{1,2,\cdots\}^M$ (such as genres of a movie) and time steps of the items $\bp^2\in\BR^M$. 
For items the user has not interacted with, this side information can have any arbitrary value. 
We can concatenate these information sources together and use $[(\hat{\bz}_{t+1}\odot \by) \oplus (\hat{\bz}_{t+1}\odot \CE(\bp^1))\oplus (\hat{\bz}_{t+1}\odot \bp^2)]$ as input to the sketching policy. 
$\CE$ is an item category embedding layer and $\oplus$ is the concatenation operator. We can define the prediction layer as, $g([j\oplus \CE(\bp^1)\oplus  \bp^2];\theta_t^{\ast})$ where the prediction is based on the item's category and a time step of adding to the sketch. %
The choice of the neural network architecture is highly flexible; the only requirement is that we need to be able to compute the gradient of the outer-level loss function w.r.t.\ ${\bz}_t$.

\subsection{Experimental Setup}
We provide additional details on the network architectures and hyper-parameters in this section.

\paragraph{Networks and Hyper-parameters.}
We use collaborative filtering based neural network $\Theta$ as the prediction module for every method. The prediction network takes user embedding $\Theta(u)$, the item index, $j$, and produces the explicit (implicit) rating on the item $j$. 
We use fixed $32$-dimensional vectors for the prior user embedding $\Theta(u)$ and the item embeddings for all datasets and all methods. The policy layer neural network $f$ consists of two hidden layers of $128$ nodes, dropout regularization \cite{dropout-1} with fixed dropout rate of 10\%, and ReLU non-linearity \cite{dlbook} for all datasets. 
We use a batch size of $256$ users for the Movielens 1M dataset and a batch size of $128$ users for all other datasets in all cases decided based on memory requirements on a single NVIDIA 2080Ti GPU. 
We keep a queue 
of $100$ past steps for the Movielens 1M and Movielens 10M datasets; for all other datasets, we keep a queue of $50$ past steps to fit our models in the GPU.

We observe all methods are fairly robust to varying level hyper-parameter settings, possibly due to the large size of the datasets and the relatively simple neural network model architectures. 
We set the number of inner optimization gradient steps to $10$ for all datasets and all methods, and tune the inner learning rate parameter $\alpha$ from $\{0.2,0.4,0.8\}$ to adapt the user embedding parameter. 
We use the stochastic gradient descent optimizer with momentum and weight decay of $0.0002$ to optimize global user parameters. We use the Adam optimizer with a weight decay of $0.0002$ to optimize item embedding parameters and the policy parameters \cite{adam}. 
We use a fixed learning rate for all datasets, set in proportion to their batch size, following \cite{linear-sgd};  we set a learning rate of $2e-5$ ($1e-4$) for the Movielens 1M dataset with a batch size of  256 and learning rate of $1e-5$ ($5e-5$) for all other datasets, with a batch size of  128, to optimize item (user) embedding parameters. 
Similarly, we tune the policy learning rate on all datasets based on their batch size; we tune learning rate from $\{2e-4, 1e-4\}$ for the Movielens 1M dataset with a batch size of 256 and learning rate from $\{1e-4, 5e-5\}$ for all other datasets, with a batch size of 128, to optimize the policy network.

\subsection{Additional Experimental Results}

\paragraph{Approximate Gradient Directions.}
We further investigate errors propagated through our gradient approximation in the DiPS method by comparing the approximations used in DiPS and DiPS@1. 
In particular, instead of the queue-based approximation, we recreate the sketch from the start with the current policy parameters $\Phi_t$ and compute the gradient of next item loss w.r.t.\ the policy parameters. 
We are interested in the gradient directions and compare the true gradient directions with the approximate gradient directions. For the Movielens 1M dataset, on average, DiPS@1 preserves, negates, and zeros out 12\%, 6\%, and 82\% of the non-zero dimensions of the true gradient. 
In contrast, DiPS, armed with a queue of intermediate sketches, preserves, negates, and zeros out 74\%, 18\%, and 8\% of the non-zero dimensions of the true gradient. 
We note that the gradient flows only through items in the intermediate sketches. 
Since keeping intermediate sketches in the queue can somewhat recreate the streaming process from start, we observe that only $0.3\%$ and $0.03\%$ of the dimensions with zero true gradients are non-zero for the gradients estimated by DiPS and DiPS@1. 
We observe a similar behavior on all other datasets. 
This observation suggests DiPS with a queue of historical sketches can pass approximately correct gradient directions without incurring the heavy computational cost of recreating the sketch at every time step.

\end{document}